\documentclass{elsart}
\usepackage[colorlinks]{hyperref}
\newcounter{bla}
\newenvironment{refnummer}{%
\list{[\arabic{bla}]}%
{\usecounter{bla}%
 \setlength{\itemindent}{0pt}%
 \setlength{\topsep}{0pt}%
 \setlength{\itemsep}{0pt}%
 \setlength{\labelsep}{2pt}%
 \setlength{\listparindent}{0pt}%
 \settowidth{\labelwidth}{[9]}%
 \setlength{\leftmargin}{\labelwidth}%
 \addtolength{\leftmargin}{\labelsep}%
 \setlength{\rightmargin}{0pt}}}
 {\endlist}

\begin{document}
\begin{frontmatter}

\title{The repository of physical models used for the CODATA-2002 FPC (re-)evaluation}

\author[a]{Andrey S. Siver},

\address[a]{Institute for High Energy Physics, Russia}

\begin{abstract}
We present PAREVAL package containing a repository of theoretical physical models used for (re-)evaluation of the fundamental physical constants (FPC). It holds all necessary data for building 105 (so called) observational equations and can be used in high precision calculations. Among repository models there are expressions for energy levels of hydrogen and deuterium (with 16 types of contributions), electron and muon magnetic moment anomalies, muonium ground-state hyperfine splitting, Zeeman energy levels in muonium. Each model is represented as {\sl Mathematica} module with XML meta-data keeping information about the model (including data on dependence from others models). There are also modules for working with the basic FPC.

\begin{flushleft}

\end{flushleft}


\end{abstract}

\end{frontmatter}


{\bf PROGRAM SUMMARY}
  
\begin{small}
\noindent
{\em Manuscript Title:} The repository of physical models used for the CODATA-2002 FPC (re-)evaluation                                       \\
{\em Authors:} Andrey S. Siver                                               \\
{\em Program Title:} PAREVAL                                          \\
{\em Journal Reference:}                                      \\
{\em Catalogue identifier:}                                   \\
{\em Licensing provisions:} none                                  \\
{\em Programming language:} {\sl Mathematica}                                  \\
{\em Computer:} any                                              \\
{\em Operating system:} any                                      \\
{\em RAM:} 10 Mbytes                                              
{\em Classification:}                                         \\
{\em Nature of problem:} PAREVAL package contains the repository of physical models (i.e. expressions arising from physical theories) which can be written in terms of the basic fundamental physical constants (FPC).\\  
PAREVAL package was designed with following ideas in mind: (1) to simplify the procedure of the addition of new physical models (`breadth evolution');  (2) to simplify the procedure of the attachment of new models built from more general principles (`depth evolution'); (3) give an environment for comparison of the models.\\
At present the models are precisely taken from [1,2] and the correctness of their representation in {\sl Mathematica} was checked during our re-evaluation of the basic FPC.\\
\\
{\em References:} 
\begin{refnummer}
\item P.~J. Mohr and B.~N.~Taylor, The 2002 CODATA Recommended Values of the Fundamental Physical Constants, {\em  Reviews of Modern Physics} {\bf 77} 1 (2005).
\item P.~J. Mohr and B.~N.~Taylor, CODATA recommended values of the fundamental physical constants: 1998, {\em Reviews of Modern Physics} {\bf 72} 351 (2000)
\end{refnummer}
\end{small}

\href{http://sirius.ihep.su/~siver/par-0.9.4f.zip}{DOWNLOAD `PAREVAL' (3.5 Mb)}

\newpage


\hspace{1pc}
{\bf LONG WRITE-UP}
\section{Introduction}
PAREVAL package is a set of {\sl Mathematica} modules and notebooks which can be arranged by following:
\begin{itemize}
\item The repository of physical models (i.e. {\sl Mathematica} modules with some meta-data in XML);
\item Modules for the FPC usage. Contain data and functions for the FPC usage in high-precision calculations;
\item Notebooks with examples. The most interesting example is probably our re-evaluation of the basic FPC-2002;
\item Module for the final presentation of correlated quantities. Contain functions for the final rounding of results (according with techniques described in \cite{pre-2}). 
\end{itemize}

Each repository module defines some {\sl Mathematica} symbols (variables or functions). Value of the defined symbol usually corresponds to some physical expression. So the content of the repository we can represent as the realization of some phenomenological physical knowledge.

Now the models are precisely taken from \cite{CODATA-2002} and the correctness of their representation was checked during our re-evaluation of the basic FPC.

Scientists from ``Fundamental Constants Data Center'' at National Institute of Standards and Technology (NIST) carried out a great work on collecting data which can be used for precision determination of the values of the FPC. We think that the theoretical expressions, on which the evaluation is based on, are interesting by itself and it's the main motivation for designing the PAREVAL package. 

New theoretical expressions can be easily add to the repository by the user. For example, one could find a module (M1), which should be improved, introduce a new {\sl Mathematica} symbol (or function) into it, place calculations for the symbol into separate module file and at last include the module name in dependence list in XML meta-data file of the first module M1 (thus it will be loaded by the corresponding repository function before the M1 module). We represent this to ourselves as `breadth evolution' of the repository. 

Since {\sl Mathematica} allows {\it in principle} to carry out {\it any} calculations, it's possible to replace a model by another module calculated from more general principles. We call this by `depth evolution'.

As one can see repository models can be written in terms of the basic fundamental physical constants. Let's notice that the basic FPC is a some subset of the FPC which have following property: each of the FPC can be precisely expressed as the function of the basic ones. So after evaluation the basic FPC all the FPC can be calculated.
\subsection{A simple technology for accounting computational uncertainties}
\label{calc-unc}
In our FPC-2002 evaluation we have been generally based on the reviews \cite{CODATA-2002}, \cite{CODATA-98}. But in the realization of the models we embedded some our ideas. We've designed a simple technology for accounting for values with calculation uncertainties. 

Some parts of theoretical expressions for physical quantities come from numerical calculations, fits or approximations and only numerical values are usually known for them. They usually have uncertainties which should be accounted for. We solve the problem as following. Instead of use a scalar value $A$ (which have an uncertainty $U$) we insert an object `ss[A,U,Label]' ({\sl Mathematica} function) into corresponding {\sl Mathematica} expression and have created several function `svalue', `suncer' which suppress the {`ss[...]'} objects (from the expression) and use corresponding information from it:
\begin{itemize}
\item `svalue[expr]' - gets out the value of the expression `expr' in which `A' value is used for every `ss[A,U,Label]';
\item `suncer[expr]' - gets out the uncertainty of the expression `expr' where objects \{`ss[A,U,Label]'\} are considered as independent stochastic quantities (with corresponding normal distributions) with uncertainties \{`U'\}.
\end{itemize}
`Label' data can be used to identify the object in the expression.
\section{The repository of physical models}
\label{repo}
Each repository model is represented as {\sl Mathematica} module (i.e. `NAME.m' file, where `NAME' is the name of the model) and `NAME.xml' file with the following structure of XML tags:
\begin{verbatim}
<m> <!-- the model opening tag -->
        <dep> `sequence of the names of models which should be loaded 
              before this model' </dep>  	
        <p> <!-- parameters opening tag -->
                <inp> `sequence of the input parameters' </inp>
                <out> `sequence of the output parameters' </out>
                <sha> `sequence of the shared parameters' </sha>
        </p> <!-- parameters closing tag -->
        <title> `the title of the model' </title>	
</m> <!-- the model closing tag -->
\end{verbatim}

The repository models are listed in the Appendix A.

These models were enough for building 105 (so called) observational equations (i.e. expressions which : (1) were measured experimentally (or in some special cases calculated or estimated theoretically); (2) can be expressed in terms of the FPC) selected by the CODATA-2002.

These observational equations can be separated into two groups "A" (principle observational equations for determination of the Rydberg constant) and "B" (principle observational equations for determination of some others basic FPC).

In order to show the correctness of the repository models we prepared several web-pages with the data on the observational equations\footnote{Web-browser should support MathML format to display the pages. For Internet Explorer you can download MathPlayer plug-in: \href{http://www.dessci.com/en/products/mathplayer/}{http://www.dessci.com/en/products/mathplayer/}.}. 

Firstly, we would like to present a table where experimental data are compared with theoretical ones: 
\begin{itemize}
\item Data on the observational equations. Group "A" \href{http://hermes.ihep.su/fpc/table-oes-a.xhtml} {http://hermes.ihep.su/fpc/table-oes-a.xhtml};
\item Data on the observational equations. Group "B" \href{http://hermes.ihep.su/fpc/table-oes-b.xhtml}{http://hermes.ihep.su/fpc/table-oes-b.xhtml};
\end{itemize}

Let's make some comment:
\begin{itemize}
\item Correlation coefficients of the input data for FPC-2002 evaluation are presented on the page: \href{http://hermes.ihep.su/fpc/table-oes-corr.xhtml} {http://hermes.ihep.su/fpc/table-oes-corr.xhtml}
\item "ID" column: The numbers after A'-, B'- indicate on the position from the beginning of the group (For "A" group it corresponds to CODATA-2002 ``LSA index'' (see, for example, a table from NIST web site \\ \href{http://physics.nist.gov/cuu/LSAData/qdata.pdf}{http://physics.nist.gov/cuu/LSAData/qdata.pdf}), "B'1" corresponds to 51 ``LSA index'', "B'2" to 52 and so on). So ordering of the observational equations is precisely the same as used in the CODATA-2002 evaluation);
\item "Exp. uncer.", "Rel. exp. unc.": experimental and relative experimental uncertainties correspondingly; 
\item "Theor. uncer.", "Rel. theor. unc.": theoretical and relative theoretical uncertainties correspondingly. Theoretical uncertainty arises from the uncertainties of the FPC (including $\delta$-corrections due to uncalculated terms) according to the law of uncertainty propagation;
\item "Calc. uncer.": Calculational uncertainty was shortly discussed in the section \ref{calc-unc}.  It's due to the uncertainties that some of auxiliary parameters in theoretical expressions have. Large calculation uncertainty for A'1 observational equation is mainly due to uncertainty of $C_{50}$ coefficient from expression for three photon correction (\cite{CODATA-2002}, p.82). We believe that it was reduced by introduction the $\rm \delta_N(n,L,j)$ \footnote{corrections to the energy levels of the hydrogen and deuterium due to uncalculated terms} with corresponding auxilary observational equations (and corresponding correlation coefficients). From other hand it shows that computational uncertainty is rough estimation of the uncertainty (but possibly more reliable);
\item "Cal.uncer/Th.uncer" : Ratio of the computational uncertainty to the theoretical one. It's is an indicator: usually it should be less then 0.1;
\item "(Exp.-Theor.)/Unc." is an indicator of agreement of theoretical and experimental values. "Unc." is the total uncertainty: ${\rm Unc.}=\sqrt{U_{experiment}^2+U_{theory}^2}$ (where $U_X$ is corresponding uncertainty); 
\item "Original expression" is original expression for observational equation from the repository model.       
\end{itemize}

Secondly, we show tables of data calculated by NIST (after FPC-2002 evaluation) and with the PAREVAL:   

\begin{itemize}
\item Comparison values of the observational equations of the group "A": CODATA-2002/NIST vs. PAREVAL\\ \href{http://hermes.ihep.su/fpc/table-oes-a_NIST_PARE.xhtml}{http://hermes.ihep.su/fpc/table-oes-a\_NIST\_PARE.xhtml};
\item Comparison values of the observational equations of the group "B": CODATA-2002/NIST vs. PAREVAL\\ \href{http://hermes.ihep.su/fpc/table-oes-b_NIST_PARE.xhtml}{http://hermes.ihep.su/fpc/table-oes-b\_NIST\_PARE.xhtml}.
\end{itemize}

Comments are followings: 
\begin{itemize}
\item Rows are sorted by the values of fifth column -- indicator of computational agreement;
\item "Exp.Uncer." is experimental uncertainty. Usually it defines the level of required precision so we compare the values via it: it should be less then 10\% normally (in our case the maximum value 
is less then 7\% with maximum for A'14 item).
\end{itemize}

So as one can see we have good agreement with NIST calculations of the observational equations.

\section{The results of our FPC-2002 re-evaluation}
The most interesting example of the repository usage is probably our re-evaluation of the basic FPC-2002.

We obtained good agreement with CODATA-2002 values of the basic FPC: the maximum value of the ${\rm `Pull'}=|(z_{PAREVAL}-z_{CODATA})/U_{PAREVAL}|$ is less then 4\% (where by $z_X$ we denote corresponding value of the basic FPC).

All the constants uncertainties obtained by us are very closely to CODATA-2002 ones in the meaning $|U_{PAREVAL}/U_{CODATA}-1|<1\%$ , where by $U_X$ we denoted uncertainty of the constant (for CODATA-2002 and our results).  

One can see results of our evaluation from the followings web-pages\footnote{Web-browser should support MathML format to display the pages. For Internet Explorer you can download MathPlayer plug-in: \href{http://www.dessci.com/en/products/mathplayer/}{http://www.dessci.com/en/products/mathplayer/}.}: 
\begin{itemize}
\item The table with the basic FPC \href{http://hermes.ihep.su/fpc/table-fpc.xhtml}{http://hermes.ihep.su/fpc/table-fpc.xhtml};
\item The correlation matrix of the basic FPC \href{http://hermes.ihep.su/fpc/table-fpc-corr.xhtml}{http://hermes.ihep.su/fpc/table-fpc-corr.xhtml}; 
\end{itemize}

To determine the number of significant figures for final rounding procedure for the uncertainties of the basic FPC we use an expression \cite{pre-2}: 
$$P_U^{th} = \left \lceil {\frac 1 2}
\log_{10}\left({\frac {n} {4 \cdot
 \lambda^C_{min}}} \right ) \right \rceil$$
, where $\lambda^C_{min}$ is the minimal eigenvalue of the correlation matrix of the basic FPC. So we got $P_U^{th}=4$ (as one can see from the table of the basic FPC from the web-page). 

In order to save correlation matrix to be positive defined after final rounding procedure we use the following expression for number of significant figures: \cite{pre-2}:
$${ A_{C}^{th} = \left \lceil \log_{10}\left({\frac{n-1}
{2 \cdot \lambda^C_{min}}}\right ) \right \rceil }.$$
We got $A_{C}^{th}=7$.

\section{Conclusion}
By means of PAREVAL package the modern theoretical expressions which were collected by ``Fundamental Constants Data Center'' at NIST become accessible for scientific community as repository of physical models in {\sl Mathematica} computer algebra system. So the repository  represents  a reference-book of standard physical expressions conjugated with calculation media. This allows to see, calculate and compare theoretical expressions with simplicity. 
\section{Acknowledgments}
Author would like to thank Zenin O. V., Ezhela V. V. and P.~J.~Mohr.

The work was not supported by the project RFFI-05-07-90191-w.
\section{Appendix A}
The repository modules are listed in following two tables\footnote{All indicated pages in the tables refer usually to \cite{CODATA-2002} if it's not written anything else.}:

\begin{tabular}{|l|l|l|}
\hline 
Title & Output Symbol(s) & Module file \\ 
\hline 
\hline 
Electron magnetic moment anomaly & ${a_{"e"}}[\alpha,{dae}]$ & a-e.m\\
(p.474-476)&& \\
\hline
Muon magnetic moment anomaly  & ${a_{"\mu "}}[\alpha,{damu}]$ & a-mu.m \\ 
(p. 86-89)& & \\
\hline
\multicolumn{3}{|c|}{Models for energy levels of hydrogen and deuterium atoms:}\\
\hline 
- General contributions (p.77-84) & {E1byParts} {E1tot} & e-l-2.m \\\hline 
- Two-photon corrections (p.80) & ${{{E1}}_4}$ & e-l\_4.m \\\hline
- Three-photon corrections (p.82) & ${{{E1}}_6}$ & e-l\_6.m \\ \hline
- Finite nuclear size contribution  	& ${{{E1}}_{{NS}}}$ & e-l\_fns.m \\ 
(p.82)					& & \\
\hline 
- Relativistic recoil non-leading  	& ${{{E1}}_{{R01}}}$ & e-l\_rr2.m \\ 
terms contribution (p.78)		& & \\ 
\hline
- Radiative-recoil correction & ${{{E1}}_{{RR}}}$ & e-l\_rr.m \\
(p.83)				& & \\
\hline
\hline
Ratios of bound-particle to	& {rgeHge} {rgpHgp} 	& g-fact-rat.m \\
free-particle g-factors (p. 93)	& {geDge} {gmupMugmup} 	& \\
				& {geMuge} {gdDgd}	& \\
\hline				
Bound-state g-factors (for bound  			
				& $\rm gg_{"C"}[\delta]$ 	& g-fact.m \\ 
electron in $\rm ^{12}C^{5+}$ and $\rm ^{16}O^{7+}$)	
				& $\rm gg_{"O"}[\delta]$		& \\
(p.89)				& ${g_{"e"}}$ & \\
\hline
\end{tabular}

\newpage
\begin{tabular}{|l|l|l|}
\hline
Title & Output Symbols & Module file \\
\hline
\hline
Muoniom ground-state hyperfine  & {dnMUth} 	& mu-hs.m \\ 
splitting (p.94-96)		& 		& \\
\hline
Zeeman energy levels in  muonium  & $\nu [{fp}]$	& mu-zee.m \\ 
(\cite{CODATA-98},p.386-387; \cite{CODATA-2002},p.94-96)	& 		& \\
\hline
Ground-state ionization energies  	& {matrix4}	& table-e\_b.m \\ 
(p. 11, table IV) 			& 		& \\
\hline
\hline
Set of the observational equations  		& {OEListDefA1}		& tdata-a.m \\ 
for data "A" for the FPC-2002			& {OEListDefDelta}	& \\
evaluation (p.57)&&\\
\hline
Set of the observational equations  	& {OElistDefB} {OElist} & tdata-b.m \\
for data "B" for the FPC-2002		& 			& \\
evaluation (p.59-60)&&\\
\hline
\hline
Mathematica variables for   		& {varsFPC} {subsFPC} & reFPC-2002a.m \\ 
calculations with the FPC-2002 		& {uncerFPC} {unitsFPC}  	& \\
(the results of our re-evaluation)	& {namesFPC} 	& \\					
					& {corrFPC} {covFPC}		& \\
					& {texsymbFPC}			& \\
\hline					
Functions for accounting for 		& svalue[expr] & nnum.m \\ 
computational uncertainties		& suncer[expr] & \\
\hline
\end{tabular}


\begin{thebibliography}{100}
\bibitem{CODATA-2002} P.J. Mohr and B.N. Taylor, The 2002 CODATA Recommended Values of the Fundamental Physical Constants, {\em  Reviews of Modern Physics} {\bf 77} 1 (2005).

\bibitem{CODATA-98} P.J. Mohr and B.N.~Taylor,
 ``CODATA recommended values of the fundamental physical constants: 1998,'' Rev. Mod. Phys. {\bf 72} 351 (2000)

\bibitem{pre-2}
  V.~V.~Ezhela, Y.~V.~Kuyanov, V.~N.~Larin, A.~S.~Siver,
  arXiv:physics/0409117 
\end{thebibliography}
\end{document}